\documentclass[12pt,draftclsnofoot,onecolumn]{IEEEtran}
\usepackage{MnSymbol}
\usepackage{amsmath}
\usepackage{lipsum}
\usepackage{optidef}
\usepackage{amsmath}
\usepackage{bm}
\usepackage{amsfonts}
\usepackage{paralist}
\usepackage{cite}
\usepackage{xcolor}
\usepackage{hyperref}
\makeatletter
\newcommand*{\rom}[1]{\expandafter\@slowromancap\romannumeral #1@}

\makeatother
\begin{document}
	\title{Securing RIS-Aided Wireless Networks Against Full Duplex Active Eavesdropping}

 \author{Atefeh Zakeri, and S. Mohammad~Razavizadeh \\ \ \\
	{ \small Mobile Broadband Network Research Group \\ School of Electrical Enginnering\\ Iran University of Science and Technology (IUST) \\
    \indent \;  \url{azakeri@iust.ac.ir} , \url{smrazavi@iust.ac.ir} }}
\normalsize
	\maketitle
\begin{abstract}
	This paper investigates the physical layer security of a Reconfigurable Intelligent Surface (RIS)-aided wireless network in the presence of full-duplex active eavesdropping. In this scenario, the RIS cooperates with the Base Station (BS) to transfer information to the intended user while an active attacker attempts to intercept the information through a wiretap channel. In addition, the attacker sends jamming signals to interfere with the legitimate user's reception of the signal and increase the eavesdropping rate.
	Our objective is to maximize the secrecy rate by jointly optimizing the active and passive beamformers at the BS and RIS, respectively. To solve the resulting non-convex optimization problem, we propose a solution that decomposes it into two disjoint beamforming design sub-problems solved iteratively using Alternating Optimization (AO) techniques.
	Numerical analysis is conducted to evaluate the impacts of varying the number of active attacking antennas and elements of the RIS on the secrecy performance of the considered systems under the presence of jamming signals sent by the attacker. The results demonstrate the importance of considering the impact of jamming signals on physical layer security in RIS-aided wireless networks. Overall, our work contributes to the growing body of literature on RIS-aided wireless networks and highlights the need to address the effects of jamming and active eavesdropping signals in such systems.
\end{abstract}
\section{Introduction}
Ensuring security is crucial in 5G and 6G wireless communication networks, as wireless networks have become an indispensable part of our daily lives for exchanging data and communicating with ease. However, malicious activities such as eavesdropping and jamming attacks can compromise the secrecy and integrity of wireless communications \cite{Ahmad2019}. The use of encryption algorithms is a common method to ensure security in wireless networks, but due to the rapid development of technology and computing power, physical layer security checks have been introduced as a supplement to conventional methods. This approach can evaluate the security performance of wireless networks using parameters like the secrecy rate, which indicates the upper limit of the effective rate that can be transmitted confidentially. Passive eavesdropping attacks involve intercepting signals without transmitting any of their own, while active eavesdropping attacks send disruptive signals to interfere with legitimate transmissions. An active attack whose sole purpose is to cause interference is known as a jamming attack \cite{Yongpeng2018,Yiliang2017}.

Reflective Intelligent Surfaces (RIS) are a fundamental technology that can improve performance of wireless communications and increase system transmission rates by adjusting reflection coefficients. RIS comprises a flat surface that consists of numerous reflective elements, each capable of inducing independent amplitude or phase shifts to the propagated signal and implementing a smart radio environment by controlling the properties of signal propagation \cite{Qingqing2021,Shimin2020}. RIS has important applications in new wireless networks as it offers various benefits including reduced interference, better coverage, and improved energy efficiency. Additionally, RIS can be used in conjunction with existing wireless systems, making it a viable option for improving the security and efficiency of both current and future wireless networks\cite{Jie2019}.

The use of RIS to improve physical layer security has been widely studied in literature. For instance, \cite{xiao2020} investigated the maximization of coverage rate between the transmitter and the receiver in the presence of an attacker when RIS is present. References \cite{Feng2021,Zhang2021} explored the use of RIS to enhance physical layer security in wireless communications against multi-antenna eavesdroppers. They introduced an effective algorithm for the joint optimization of active and passive beamforming. In reference \cite{Liu2021}, a secure wireless communication system using Multi-Input Multi-Output (MIMO) based on RIS was proposed. The Base Station (BS) is equipped with multiple antennas to communicate with a legitimate multi-antenna user while protecting against multi-antenna passive eavesdroppers. They proposed an alternating optimization algorithm with the Taylor series expansion method and predicted gradient ascent method to design the transfer covariance matrix to maximize the secrecy rate.  Authors in \cite{Zheng2020} examined the impact of RIS on secure wireless transmission, whereby an RIS is deployed to assist the secure MIMO system to improve privacy performance. Artificial Noise (AN) is employed to mislead the passive eavesdropper. Reference \cite{Hehao2021} investigated the advantage of using RIS in multi-user Multi-Input Single-Output (MISO) systems in the presence of passive eavesdroppers. They showed that the secrecy rate could be maximized by co-designing secure beamforming, AN, and RIS phase shift. An iterative optimization method was proposed to deal with the formulated non-convex problem.  Reference \cite{Yang2021} proposed a RIS-assisted anti-jamming strategy for wireless communication security. Their goal was to maximize the system's rate in the presence of a smart jammer, where the jammer tries to spoil the quality of the desired transmissions by jamming the signal on the channels of legitimate users. In addition, reference \cite{Sai2021} considered a RIS-based backscatter communication system against jamming attack. Here, an attacker attempts to prevent the legitimate user from receiving the desired signal by emitting a jamming signal. To resist the jamming attack, an RIS is deployed near the user and acts as a transmitter to convert all the received signals into the desired signal. The objective of this paper is to maximize the Signal-to-Interference-plus-Noise Ratio (SINR) at the user, subject to power constraints at the source. In reference \cite{Tang2021}, an aerial RIS was proposed, whereby the impact of jamming attacks can be reduced by increasing the legal signal and transmission rate. Conversely, in reference \cite{Bin2020}, RIS was used as a jammer to attack legal communication, without using any internal energy to generate jamming signals, which minimizes the received signal power in the legitimate receiver. To achieve secure transmission, RIS is used as a jamming device to create and broadcast jamming signals, thereby disturbing the reception of eavesdroppers, as shown in references \cite{Majid2021,Sai2022}. In reference \cite{Wang2020}, a co-jammer was introduced in the presence of RIS, who tries to mislead the eavesdroppers. This increases the secrecy rate and effective energy by jamming signal transmission.

Full-duplex  communication is a promising method that supports simultaneous transmission and reception of radio signals on the same frequency. Therefore, it can significantly improve spectrum efficiency and reduce communication delay in future wireless networks \cite{David2017,Tingjun2021}. In a communication system based on RIS, to improve performance, reference  \cite{XiaoTang2021,Khalid2021} considered a Full-duplex legal receiver that sends the jamming signal to the receiver to mitigate the eavesdropper. This leads to the joint optimization of received beamforming, signal jamming, and passive beamforming, thereby seeking to increase the security rate.
As discussed above, in most of the previous papers on RIS assisted networks, only passive or active attacks have been considered.

Motivated by the need for improved wireless network security, this paper examines the physical layer security of a RIS-aided wireless network in the presence of a full-duplex active attacker. The RIS and BS cooperate to transfer information to the intended user while the attacker intercepts the information through a wiretap channel and sends jamming signals to interfere with the legitimate user's reception of the signal. The objective is to maximize secrecy rate through joint optimization of active and passive beamformers at the BS and RIS, respectively. To solve this non-convex optimization problem, the paper proposes an iterative solution using Alternating Optimization (AO) techniques. Numerical analysis demonstrates the importance of considering the impact of jamming signals on physical layer security in RIS-aided wireless networks. Overall, this work contributes to the growing literature on RIS-aided wireless networks and highlights the need to address the effects of jamming signals in such systems.
The paper presents several contributions, including:
\begin{itemize}
 	\item Investigating the security problem of the physical layer in a system based on RIS in the presence of a full-duplex active attacker who can act as an eavesdropper and jammer simultaneously.
 	
 	\item Including direct channel performance evaluation assuming low probability of disconnection between the BS and user, unlike many previous studies that ignore this aspect.
 	
 	\item Formulating a non-convex optimization problem for secrecy rate maximization by jointly optimizing beamforming and reflection coefficients for the RIS network. An alternative algorithm is proposed to solve the problem as two separate sub-problems.
 	
 	\item Demonstrating through simulation results the significant impact of a jammer on secrecy rate and emphasizing the importance of RIS in improving it. The effect of changing the number of active attacking antennas is also analyzed, showing an increase in secrecy rate with an increase in the number of active attacking antennas.
 	
 \end{itemize}
 The rest of this paper is organized as follows. The system model is discussed in Section \ref{system model}. In Section \ref{proposed method}, the achievable secrecy rate, the problem formulation and the proposed algorithm for solving the optimization problem is proposed. The simulation results are provided in Section \ref{numerical results}. Finally, concluding remarks are given in Section \ref{conclusion}.


\section{System model} \label{system model}
\begin{figure}[t]
	\centering
	\includegraphics[scale=0.9]{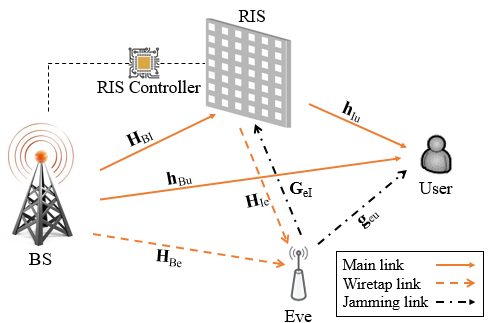}
	\renewcommand{\figurename}{Fig.}
	\caption{System Model of the RIS-assisted Communication Network Against an Active Eavesdropper.}
	
	\label{fig.system}
\end{figure}

In this paper, we consider a RIS-assisted communication system, illustrated in \hyperref[fig.system]{Fig.1}, where a base station with $K$ antennas serves a single-antenna user in the presence of an active eavesdropper (denoted as "eve"). The eavesdropper is equipped with $ N_r $ receiving antennas and $ N_t $ transmitting antennas, and is assumed to have perfect self-interference cancellation, as described in \cite{Jung2010,Sabharwal2014}. The RIS, consisting of $ L $ reflecting elements, cooperates with the base station to transmit information to the user by changing the phase of the signal received from the base station and retransmitting it. However, the full-duplex active eavesdropper listens to the information through the wiretap channel and sends a jamming signal to the RIS and the user to disrupt the legitimate user's reception of the signal and decrease the secrecy rate.

The channel gains from BS to the RIS, BS to the user, BS to the active eavesdropper, RIS to the user, RIS to the active eavesdropper are denoted as $ \mathbf{H}_{BI}\in\mathbb{C}^{L \times K}  $ , $\mathbf{h}_{Bu}\in\mathbb{C}^{K\times 1}$, $\mathbf{H}_{Be}\in\mathbb{C}^{N_{r}\times K}$, $\mathbf{h}_{Iu}\in\mathbb{C}^{L\times 1}$ and $\mathbf{H}_{Ie}\in\mathbb{C}^{N_{r}\times L}$ respectively. The baseband equivalent channels from eavesdropper to RIS and eavesdropper to user are denoted by $\mathbf{G}_{eI}\in\mathbb{C}^{L\times N_{t}}$ and $\mathbf{g}_{eu}\in\mathbb{C}^{N_{t}\times 1}$.
In this paper, we assume that the Channel State Information (CSI) of all the channels are perfectly known. This can be achieved
methods such as the local oscillator power leakage from the eavesdropper receivers’ RF frontend \cite{Mukherjee2012} or eavesdropper can also be an active user in the secure transmission system but untrusted by user \cite{Feng2021}. The received signals at the legitimate user and eavesdropper are given as
\begin{align} \label{eq.signals}
	Y_{u}&=\mathbf{h}_{Bu}^{H}\mathbf{w}s+\mathbf{g}_{eu}^{H}\mathbf{v}a+\mathbf{h}_{Iu}^{H}\mathbf{\mathbf{\theta}}\left(\mathbf{H}_{BI}\mathbf{w}s+\mathbf{G}_{eI}\mathbf{v}a\right)+n_{u} &&\\
	Y_{e}&=\mathbf{H}_{Be}\mathbf{w}s+\mathbf{H}_{Ie}\mathbf{\theta}(\mathbf{H}_{BI}\mathbf{w}s+\mathbf{G}_{eI}\mathbf{v}a)+n_{e} &&
\end{align}
we denote $\theta=diag(B_{1}e^{ja_{1}}, B_{2}e^{ja_{2}},...,B_{L}e^{ja_{L}})$ as the reflection coefficient  matrix of the RIS. $a_{l}$ and $B_{l}$ with $l=[1,2,...,L]$ are the phase shift and amplitude at the $l$th RIS element. The signal transmitted from the BS is given by $\mathbf{x}=\mathbf{w}s$, where $\mathbf{w}$ and $s\sim CN(0,1)$ denote beamforming vector and information bearing for user, respectively. Moreover, $\mathbf{v}$ representing the beamforming vector at the eavesdropper is used to transmit the jamming signal $a$. where $n_{u}$ and $n_{e}\sim CN(0,\sigma ^{2})$ are the complex Additive White Gaussian Noise (AWGN) related to the user and eavesdropper. The achievable rates at the legitimate user and the active eavesdropper are given by
\begin{align} \label{eq.rates}
	R_{u}&=log\left(1+\frac{\left|(\mathbf{h}_{Bu}^H+\mathbf{h}_{Iu}^H \theta \mathbf{H}_{BI})\mathbf{w}\right|^{2}}{\left|(\mathbf{g}_{eu}^{H}+\mathbf{h}_{Iu}^H\theta \mathbf{G}_{eI})\mathbf{v}\right|^{2}+\sigma_{u}^{2}}\right) &&\\
	R_{e}&=log\left(1+\frac{\left\|\big(\mathbf{H}_{Be}+\mathbf{H}_{Ie}\theta \mathbf{H}_{BI}\big)\mathbf{w}\right\|^{2}}{\left\|\big(\mathbf{H}_{Ie}\theta \mathbf{G}_{eI}\big)\mathbf{v}\right\|^{2}+\sigma_{e}^{2}}\right) &&
\end{align}
\\Thus the achievable secrecy rate can be written as
\begin{align} \label{eq.secracy rate}
R_{s}=[R_{u}-R_{e}]^{+}&&
\end{align}
Where $[z]^+ =max(z, 0)$, as the cases with a non-positive secrecy rate is lacking meaning in this context, for the discussions henceforth we delete the $[.]^+$ operation \cite{Miao2019}.


\section{Beamforming Design} \label{proposed method}
Based on the aforementioned discussion, this paper focuses on maximizing the secrecy rate by jointly optimizing the transmit beamforming vector $\mathbf{w}$ at the BS and the phase shifts $\mathbf{\theta}$ at the RIS. This problem can be expressed as
\begin{subequations}  \label{eq.optimize}
\begin{align}
	\max\limits_{w,\theta} \quad &R_{s} &&\\
	s.t.\quad &\|\mathbf{w}\|_{2}\leq P_{BS} &&\\
	&|\theta_l|=1,\quad l\in[1,2,...,L] &&
\end{align}
\end{subequations}
Evidently, this is a Non-deterministic Polynomial-time hard (NP-hard) problem due to the non-convex objective function and unit modulus constraints. Generally, it is impossible to find the optimal solution directly using existed algorithms. Thus, in this paper, we propose a low-complexity AO based algorithm to solve this. In this way, we turn the optimization problem into two separate sub-problems and examine each of the created sub-problems separately. In the following, a method to solve these sub-problems will be discussed.


\subsection{Active Beamforming}
First, we assume that the parameter $\theta$ are fixed and derive the optimal value of $\mathbf{w}$. To this end, the objective function in problem \eqref{eq.optimize} is reformulated. Therefore, sub-problem  will be
\begin{subequations}  \label{eq.active}
\begin{align}
	\max\limits_{w} \quad &R_{s} &&\\
	s.t.\quad &\|\mathbf{w}\|_{2}\leq P_{BS} &&
\end{align}
\end{subequations}
The objective function of this problem is non-convex. By defining $\mathbf{W}=\mathbf{w}\mathbf{w}^{H}$ and $\mathbf{V}=\mathbf{v}\mathbf{v}^{H}$ To facilitate the discussions, we rearrange the terms within the rate as
\begin{align} \label{eq.rate Active}
R_{u}&= log\left(I+(\mathbf{H}_{Bu}\mathbf{W}\mathbf{H}_{Bu}^{H})(I+\mathbf{G}_{eu}\mathbf{V}\mathbf{G}_{eu}^{H})^{-1}\right)&& \\
R_{e}&= log\left(I+(\mathbf{H}_{BeI}\mathbf{W}\mathbf{H}_{BeI}^{H})(I+\mathbf{H}_{IeI}\mathbf{V}\mathbf{H}_{IeI}^{H})^{-1}\right)&&
\end{align}
where $\mathbf{H}_{Bu}\!=\!\frac{1}{\sigma_{u}}(\mathbf{h}_{Bu}^{H}+\mathbf{h}_{Iu}^{H}\theta \mathbf{H}_{BI}), \mathbf{G}_{eu}=\frac{1}{\sigma_{u}}(\mathbf{g}_{eu}^{H}+\mathbf{h}_{Iu}^{H}\theta \mathbf{G}_{eI}), \mathbf{H}_{BeI}=\frac{1}{\sigma_{e}}(\mathbf{H}_{Be}+\mathbf{H}_{Ie}\theta \mathbf{H}_{BI})$ and $ \mathbf{H}_{Bu}=\frac{1}{\sigma_{e}}(\mathbf{H}_{Ie}\theta \mathbf{G}_{eI})$. Then, problem \eqref{eq.active} is transformed to the following problem
\begin{subequations}  \label{eq.active reform}
\begin{align}
	\max\limits_{\mathbf{w}} \quad &\underbrace{log\left(1+(\mathbf{H}_{Bu}\mathbf{w}\mathbf{w}^{H}\mathbf{H}_{Bu}^{H})(1+\mathbf{G}_{eu}\mathbf{v}\mathbf{v}^{H}\mathbf{G}_{eu}^{H})^{-1}\right)}\limits_{A_{1}}+&& \notag \\ &\underbrace{log(\mathbf{H}_{IeI}\mathbf{v}\mathbf{v}^{H}\mathbf{H}_{IeI}^{H})}\limits_{A_{2}}-&& \\  &\underbrace{log\left((\mathbf{H}_{IeI}\mathbf{v}\mathbf{v}^{H}\mathbf{H}_{IeI}^{H})+(\mathbf{H}_{BeI}\mathbf{w}\mathbf{w}^{H}\mathbf{H}_{BeI}^{H})\right)}\limits_{A_{3}}&& \notag \\
	s.t.\quad &\|\mathbf{w}\|_{2}\leq P_{BS}&&
\end{align}
\end{subequations}
Problem \eqref{eq.active reform} is still non-convex and intractable. Therefore, we employ the idea of Weighted Minimum Mean Square
Error (WMMSE) to transform the objective function into an equivalent counterpart which can be manipulated alliteratively
via the BCD method\cite{Shi2011}. To proceed, we introduce the auxiliary matrices $(\varepsilon _{i}, (i \in {1, 2, 3}), x_{j}, (j \in\ {1, 2})) $ to reformulate $A_{1}, A_{2}$ and $A_{3}$ in the objective function in problem, respectively. First, let us consider the Mean Square Error (MSE) matrix function of $A_{1}$ as follows:
\begin{align} \label{eq.wmmse1}
E_{1}(x_{1},w)= &(I-x_{1}^{H}\mathbf{H}_{Bu}\mathbf{w})(I-x_{1}^{H}\mathbf{H}_{Bu}\mathbf{w})^{H}+ && \notag\\
&x_{1}^{H}(I+\mathbf{G}_{eu}\mathbf{v}\mathbf{v}_{H}\mathbf{G}_{eu})x_{1} &&
\end{align}
Similarly, $A_{2}$ is given by
\begin{align}  \label{eq.wmmse2}
E_{2}(x_{2},w)= (I-x_{2}^{H}\mathbf{H}_{IeI}\mathbf{v})(I-x_{2}^{H}\mathbf{H}_{IeI}\mathbf{v})^{H}+x_{2}^{H}x_{2} &&
\end{align}
To solve this problem, we will need the following lemma:\\
$Lemma \; 1:$ \cite{Jose2011} Denoting $\mathbf{E}\in\mathbb{C}^{d\times d}$ as any positive definite matrix, we have the following function
\begin{align}  \label{eq.lemma}
-log(\mathbf{E})=\max \limits_{S\in\mathbb{C}^{d\times d}, S\geq 0} \quad \delta (S)
\end{align}
where $\delta (S)=-Tr(SE)+log| S| +N$. Then, the optimal solution to problem \eqref{eq.lemma} can be expressed as $S^{opt}=E^{-1}$.
\\Based on $Lemma 1$, we can obtain the following equalities,
\begin{align}  \label{eq.reform lemma}
A_{1}&=\max \limits_{\varepsilon _{1}>0,x_{1}} \; log(\varepsilon _{1}) - Tr\big[\varepsilon _{1} \big( E_{1}(x_{1},\mathbf{w})\big)\big]&& \\
A_{2}&=\max \limits_{\varepsilon _{2}>0,x_{2}} \; log(\varepsilon _{2}) - Tr\big[\varepsilon _{2} \big( E_{2}(x_{2},\mathbf{w})\big)\big]&& \\
A_{3}&=\max \limits_{\varepsilon _{3}>0} \; log(\varepsilon _{3}) - Tr\big[\varepsilon_{3}\big((\mathbf{H}_{IeI}\mathbf{v}\mathbf{v}^{H}\mathbf{H}_{IeI}^{H})+(\mathbf{H}_{BeI}\mathbf{w}\mathbf{w}^{H}\mathbf{H}_{BeI}^{H})\big)\big]&&
\end{align}
We substitute the formulas above  into problem \eqref{eq.active reform} , which is equivalently rewritten as
\begin{subequations}  \label{eq.opt active}
\begin{align}
	\max\limits_{\Omega} \quad &log(\varepsilon_{1})-Tr\big[\varepsilon_{1}\big((I-x_{1}^{H}\mathbf{H}_{Bu}\mathbf{w})(I-x_{1}^{H}\mathbf{H}_{Bu}\mathbf{w})^{H} &&\notag\\
	&+x_{1}^{H}(I+\mathbf{G}_{eu}\mathbf{v}\mathbf{v}^{H}\mathbf{G}_{eu}^{H})x_{1}\big)\big] +log(\varepsilon_{2}) &&\\
	&-Tr\big[\varepsilon_{2}\big((I-x_{2}^{H}\mathbf{H}_{IeI}\mathbf{v})(I-x_{2}^{H}\mathbf{H}_{IeI}\mathbf{v})^{H}+x^{H}_{2}x_{2}\big)\big] &&\notag\\
	&+log(\varepsilon_{3})
	-Tr\big[\varepsilon_{3}\big((\mathbf{H}_{IeI}\mathbf{v}\mathbf{v}^{H}\mathbf{H}_{IeI}^{H})+(\mathbf{H}_{BeI}\mathbf{w}\mathbf{w}^{H}\mathbf{H}_{BeI}^{H})\big)\big] &&\notag\\
	s.t.\quad&\|\mathbf{w}\|^{2}\leq P_{BS} &&\\
	&\Omega=\{\varepsilon_{1},\varepsilon_{2},\varepsilon_{3}>0,\; x_{1},x_{2}, \; \mathbf{w}\}&&
\end{align}
\end{subequations}
Then, we use the BCD algorithm and to separate problem \eqref{eq.opt active} into three sub-problem. In the sequel,  We first solve problem \eqref{eq.opt active} to optimize $x_{1},x_{2}$, given $\mathbf{w} $ and $\varepsilon_{1},\varepsilon_{2},\varepsilon_{3}$.
\begin{align} \label{eq.active x}
x_{1}&= \arg\min\limits_{x_{1}} \quad Tr[\varepsilon_{1}E_{1}(x_{1},\mathbf{w})] &&\\
x_{2}&= \arg\min\limits_{x_{2}} \quad Tr[\varepsilon_{2}E_{2}(x_{2},\mathbf{w})] &&
\end{align}
In order to solve problems above, we take into consideration their own first-order derivative, respectively, and the
closed-form solution of $x_{1}$ and $x_{2}$ is given by
\begin{align} \label{eq.solve x}
x_{1}&=(I+\mathbf{G}_{eu}\mathbf{v}\mathbf{v}^{H}\mathbf{G}_{eu}^{H}+\mathbf{H}_{Bu}\mathbf{w}\mathbf{w}^{H}\mathbf{H}_{Bu}^{H})^{-1}\mathbf{H}_{Bu}\mathbf{w} &&\\
x_{2}&=(I+\mathbf{H}_{IeI}\mathbf{v}\mathbf{v}^{H}\mathbf{H}_{IeI}^{H})^{-1}\mathbf{H}_{IeI}\mathbf{v} &&
\end{align}
In the next step, we solve problem \eqref{eq.opt active} to optimize $\varepsilon_{1},\varepsilon_{2},\varepsilon_{3}$, given $ \mathbf{w}$ and  $x_{1},x_{2}$.  It is observed that the matrices  $\varepsilon_{1},\varepsilon_{2},\varepsilon_{3}$ are independent of each other in the objective function of problem \eqref{eq.opt active}. Thus, by exploiting $Lemma 1$, the closed-form solutions of $\varepsilon_{1},\varepsilon_{2},\varepsilon_{3}$ are derived as
\begin{align} \label{eq.solve ep}
\varepsilon_{1}&=\big[(I-x_{1}^{H}\mathbf{H}_{Bu}\mathbf{w})(I-x_{1}^{H}\mathbf{H}_{Bu}\mathbf{w})^H+x_{1}^{H}(I+\mathbf{G}_{eu}\mathbf{v}\mathbf{v}^{H}\mathbf{G}_{eu}^{H})x_{1}\big]^{-1} &&\\
\varepsilon_{2}&=\big[(I-x_{2}^{H}\mathbf{H}_{IeI}\mathbf{v})(I-x_{2}^{H}\mathbf{H}_{IeI}\mathbf{v})^H+x_{2}^{H}x_{2}\big]^{-1} &&\\
\varepsilon_{3}&=\big[I+\mathbf{H}_{IeI}\mathbf{v}\mathbf{v}^{H}\mathbf{H}_{IeI}^{H}+ \mathbf{H}_{BeI}\mathbf{w}\mathbf{w}^{H}\mathbf{H}_{BeI}^{H}\big]^{-1} &&
\end{align}
Then, we solve problem \eqref{eq.opt active} to optimally design $\mathbf{w}$ given $x_{1},x_{2}$ and $\varepsilon_{1},\varepsilon_{2},\varepsilon_{3}$. To proceed, problem \eqref{eq.opt active} is equivalently rewritten with respect to $\mathbf{w}$, as
\begin{subequations}  \label{eq.Active min}
\begin{align}
	\min\limits_{w}\quad &-log(\varepsilon_{1})+Tr \big[\varepsilon_{1}\big((I-x_{1}^{H}\mathbf{H}_{Bu}\mathbf{w})(I-x_{1}^{H}\mathbf{H}_{Bu}\mathbf{w})^{H} &&\notag\\
	&+x_{1}^{H}(I+\mathbf{G}_{eu}\mathbf{v}\mathbf{v}^{H}\mathbf{G}_{eu}^{H})x_{1}\big)\big]-log(\varepsilon_{2})&&\\
	&+Tr\big[\varepsilon_{2}\big((I-x_{2}^{H}\mathbf{H}_{IeI}\mathbf{v})(I-x_{2}^{H}\mathbf{H}_{IeI}\mathbf{v})^{H}+x_{2}^{H}x_{2}\big)\big]&&\notag\\
	&-log(\varepsilon_{3})+Tr \big[\varepsilon_{3}\big((\mathbf{H}_{IeI}\mathbf{v}\mathbf{v}^{H}\mathbf{H}_{IeI}^{H})+(\mathbf{H}_{BeI}\mathbf{w}\mathbf{w}^{H}\mathbf{H}_{BeI}^{H})\big)\big] &&\notag\\
	s.t.\quad &\|\mathbf{w}\|^{2}\leq P_{BS}
\end{align}
\end{subequations}
Here we are looking for $\mathbf{w}$, so we assume other values to be constant and ignore them. After simplification, we reach the following final problem:
\begin{subequations} \label{eq.active finish}
\begin{align}
	\min\limits_{w}\quad &Tr(\varepsilon_{1}x_{1}^{H}\mathbf{H}_{Bu}\mathbf{w}\mathbf{w}^{H}\mathbf{H}_{Bu}^{H}x_{1})-Tr (\varepsilon_{1}x_{1}^{H}\mathbf{H}_{Bu}\mathbf{w})&&\notag\\
	&-Tr (\varepsilon_{1}\mathbf{w}^{H}\mathbf{H}_{Bu}^{H}x_{1})+Tr(\varepsilon_{3}\mathbf{H}_{BeI}\mathbf{w}\mathbf{w}^{H}\mathbf{H}_{BeI}^{H})&&\\
	s.t. \quad &\|\mathbf{w}\|^{2}\leq P_{BS}&&
\end{align}
\end{subequations}
The objective function of the equivalent main problem is a linear and convex function that can be solved by CVX toolbox in MATLAB.


\subsection{Passive Beamforming}
In this subsection, we design the passive beamforming at the RIS while fixing the transmit beamforming at the BS .To this end, problem \eqref{eq.optimize} is reformulated. Therefore, the sub-problem  will be
\begin{subequations} \label{eq.pasive}
\begin{align}
	\max\limits_{\theta}\quad    &R_{s} &&\\
	s.t.\quad  &|\theta_{l}|=1,\quad l\in[1,2,...,L] &&
\end{align}
\end{subequations}
The objective and constraint functions of the problem \eqref{eq.pasive} are non-convex. To facilitate the discussions, we rearrange the terms within the transmission model as
\begin{subequations}  \label{eq.pasive solve1 }
\begin{align}
	(\mathbf{h}_{Bu}^{H}+\mathbf{h}_{Iu}^{H}\theta \mathbf{H}_{BI})\mathbf{w}&=(\mathbf{h}_{Bu}^{H}\mathbf{w})+\mathbf{h}_{Iu}^{H}diag(\varphi)\mathbf{H}_{BI}\mathbf{w} &&\notag\\
	&=\mathbf{h}_{Bu}^{H}+\mathbf{h}_{Iu}^{H}diag(\mathbf{H}_{BI}\mathbf{w})\varphi &&
\end{align}
\end{subequations}
Then we consider changing the following variables.
\begin{subequations}  \label{eq.pasive solve2 }
\begin{align}
	&\big[\mathbf{h}_{Bu}^{H}\mathbf{w}, \mathbf{h}_{Iu}^{H}diag(\mathbf{H}_{BI}\mathbf{w})\big]=\mathbf{H}_{BIu} &&\\
	&[1, \varphi]^{H}={\varphi} &&
\end{align}
\end{subequations}
By applying these relations, we will be
\begin{align} \label{eq.pasive solve3}
\big\|  (\mathbf{h}_{Bu}^{H}+\mathbf{h}_{Iu}^{H}\theta \mathbf{H}_{BI})\mathbf{w}\big\|^2={\varphi}^{H}\mathbf{H}_{BIu}^{H}\mathbf{H}_{BIu}{\varphi}=Tr(\mathbf{E}_{BID}\Theta)&&
\end{align}
where $\Theta={\varphi} \; {\varphi}^{H}$. Similarly, we have
\begin{subequations}  \label{eq.pasive solve4}
\begin{align}
	&\big[\mathbf{g}_{eu}^{H}\mathbf{v}, \mathbf{h}_{Iu}^{H}diag(\mathbf{G}_{eI}\mathbf{v})\big]\big[1,\varphi\big]^{H}=\mathbf{G}_{eIu} {\varphi} &&\\
	&\big\| (\mathbf{g}_{eu}^{H}+\mathbf{h}_{Iu}^{H}\theta \mathbf{G}_{eIu})\mathbf{v} \big\|^{2}={\varphi}^{H}\mathbf{G}_{eIu}^{H}\mathbf{G}_{eIu}{\varphi}=Tr(\mathbf{E}_{eIu}\Theta)&&
\end{align}
\end{subequations}
\vspace{-0.5cm}
\begin{subequations}  \label{eq.pasive solve5}
\begin{align}
	&\big[\mathbf{H}_{Be}\mathbf{w}, \mathbf{H}_{Ie}diag(\mathbf{H}_{Be}\mathbf{w})\big]\big[1,\varphi\big]^{H}=\mathbf{H}_{BIe} {\varphi} &&\\
	&\big\| (\mathbf{H}_{Be}+\mathbf{H}_{Ie}\theta \mathbf{H}_{BI})\mathbf{w} \big\|^{2}={\varphi}^{H}\mathbf{H}_{BIe}^{H}\mathbf{H}_{BIe}{\varphi}=Tr(\mathbf{E}_{BIe}\Theta) &&
\end{align}
\end{subequations}
and finally
\begin{subequations} \label{eq.pasive solve6}
\begin{align}
	&\big[0, \mathbf{H}_{Ie}diag(\mathbf{G}_{eI}\mathbf{v}\big]\big[1,\varphi\big]^{H}=\mathbf{G}_{eIeI} {\varphi} &&\\
	&\big\| (\mathbf{H}_{Ie}\theta \mathbf{G}_{eI})\mathbf{v} \big\|^{2}={\varphi} ^{H}\mathbf{G}_{eIeI}^{H}\mathbf{G}_{eIeI}{\varphi} =Tr(\mathbf{E}_{eIeI}\Theta)&&
\end{align}
\end{subequations}
With the above definitions, the secrecy rate is reinterpreted as
\begin{align}   \label{eq.pasive rate} R_{s}=&log\left(1+\frac{Tr(\mathbf{E}_{BID}\Theta)}{Tr(\mathbf{E}_{eIu}\Theta)+\sigma_{u}^{2}}\right)-log\left(1+\frac{Tr(\mathbf{E}_{BIe}\Theta)}{Tr(\mathbf{E}_{eIeI}\Theta)
	+\sigma_{e}^{2}}\right)= &&\notag\\
&log\big(Tr(\mathbf{E}_{eIu}\Theta)+\sigma_{u}^{2}+Tr(\mathbf{E}_{BID}\Theta)\big) &&\notag\\
&-log\big(Tr(\mathbf{E}_{eIeI}\Theta)+\sigma_{e}^{2}+Tr(\mathbf{E}_{BIe}\Theta)\big) &&\notag\\
&-log\big(Tr(\mathbf{E}_{eIu}\Theta)+\sigma_{u}^2\big)+log\big(Tr(\mathbf{E}_{eIeI}\Theta)+\sigma_{e}^2\big)  &&
\end{align}
By using the same technique in \eqref{eq.reform lemma}, we can obtain the following equalities by introducing auxiliary variables $\varepsilon_{4}$ and $\varepsilon_{5}$ 
\begin{align}   \label{eq.pasive ep}
&-log\big(Tr(\mathbf{E}_{eIeI}\Theta)+\sigma_{e}^{2}+Tr(\mathbf{E}_{BIe}\Theta)\big)= && \notag\\
&\max\limits_{\varepsilon_{4}>0}\{-\varepsilon_{4} \big(Tr(\mathbf{E}_{eIeI}\Theta)+\sigma_{e}^{2}+Tr(\mathbf{E}_{BIe}\Theta)\big) + log(\varepsilon_{4}) \} &&
\end{align}
\vspace{-0.5cm}
\begin{align}  \label{eq.pasive ep2}
&-log\big(Tr(\mathbf{E}_{eIu}\Theta)+\sigma_{u}^2)\big)=  && \notag\\
&\max\limits_{\varepsilon_{5}>0}\{-\varepsilon_{5} \big(Tr(\mathbf{E}_{eIu}\Theta)+\sigma_{u}^2)\big) + log(\varepsilon_{5}) \} &&
\end{align}
With the reformulations above, the secrecy rate optimization problem is reorganized as
\begin{subequations}  \label{eq.pasive final}
\begin{align}
	\min\limits_{\varepsilon_{4},\varepsilon_{5},\Theta} \quad  R_{s}&= -log(Tr(\mathbf{E}_{eIu}\Theta)+\sigma_{u}^{2}+Tr(\mathbf{E}_{BIu}\Theta)+\sigma_{u}^{2}) && \notag\\
	&+\varepsilon_{4} (Tr(\mathbf{E}_{eIeI}\Theta)+\sigma_{e}^{2}+Tr(\mathbf{E}_{BIe}\Theta)+\sigma_{e}^{2}) &&\notag\\
	&-log(Tr(\mathbf{E}_{eIeI}\Theta)+\sigma_{e}^{2})+\varepsilon_{5} (Tr(\mathbf{E}_{eIu}\Theta)+\sigma_{u}^2) &&\notag\\
	&-log(\varepsilon_{4})- log(\varepsilon_{5}) &&\\
	s.t. \qquad &\varepsilon_{4}>0,\; \varepsilon_{5}>0 &&\\
	&\Theta\geq 0 , \quad rank(\Theta)=1 &&\\
	&\Theta_{l,l}=1, \quad \forall l\in L &&
\end{align}
\end{subequations}
For problem \eqref{eq.pasive final}, the optimized $\varepsilon_{4}$ and $\varepsilon_{5}$ is obtained:
\begin{align}   \label{eq. pasive ep4}
\varepsilon_{4}&=\frac{1}{Tr(\mathbf{E}_{eIeI}\Theta)+Tr(\mathbf{E}_{BIe}\Theta+\sigma_{e}^{2})} &&\\
\varepsilon_{5}&=\frac{1}{Tr(\mathbf{E}_{eIu}\Theta)+\sigma_{u}^2} &&
\end{align}
For the phase shift, we can resort to semidefinite relaxation and ignore the rank-1 constraint and solve the problem with
toolboxes like CVX. Then, the Gaussian randomization can be applied if the rank-1 constraint is not satisfied at the obtained optimum\cite{Hao2021,XiaoTang2021}.


\subsection{Algorithm Design}
At this point in the process, according to the two defined sub-problems, the general algorithm presented for solving the combined problem of RIS phase shift optimization and beamforming vectors is shown in \hyperref[Algorithm Design]{Algorithm 1} . The inputs of this algorithm are channel parameters and $\epsilon$, which is the maximum acceptable relative error for the minimum user security rate. In the $t$ th iteration of this algorithm, the active beamforming sub-problem is solved by using the values obtained in the previous iteration of the algorithm for the RIS phase shift in the passive beamforming sub-problem, and its answer is used as the required values to solve the passive beamforming sub-problem. This process continues until the relative error related to the minimum security rate of the user is less than $\epsilon$.

\begin{center}
\begin{tabular}{l}
\hline
\textbf{Algorithm1}  alternating iterative algorithm for solving security rate \\
\hline
\textbf{Input}: channel parameters, $\epsilon$ \\
\textbf{1)\; Initialization}: $t =0$, set $\mathbf{w}^{t}$, $\theta ^{t}$.\\
\textbf{2) \; Repeat}\\
\quad\textbf{a)} \; Given $\theta^{t}$,$ \mathbf{w}^{t}$ obtain $\varepsilon_{1,2,3}^{t+1} ,x^{t+1}$\\
\quad\textbf{b)} \; Given $\theta^{t},\varepsilon_{1,2,3}^{t+1}, x^{t+1} $ obtain $\mathbf{w}^{t+1}$\\
\quad\textbf{c)} \; Given $\theta^{t},\varepsilon_{1,2,3}^{t+1}, x^{t+1} , \mathbf{w}^{t+1}$ obtain $\varepsilon_{4,5}^{t+1}$\\
\quad\textbf{d)} \; Given $\varepsilon_{1,2,3}^{t+1}, x^{t+1} , \mathbf{w}^{t+1}, \varepsilon_{4,5}^{t+1}$ obtain $\theta^{t+1}$\\
\textbf{3)\; Until} $\frac{| R_{s}^{t}-R_{s}^{t-1}|}{R_{s}^{t-1}}\leq \epsilon$\\
\textbf{Output} Optimal transmit beamforming, passive beamforming.\\
\hline
\label{Algorithm Design}
\end{tabular}
\end{center}

Moreover, we analyze the computation complexity of the proposed algorithm. Considering the computational complexity of the sub-problem related to the active beamforming, which is defined as $\mathcal{O}\big(K^{2}+2N_{r}^{3}\big) $ and the sub-problem related to the passive beamforming, while it reaches convergence in $T_{1} $ iteration, it is defined as $\mathcal{O}\big(T_{1}(L+1)^{4.5}\big)$. The computational complexity of the main algorithm in each iteration is equal to the sum of their complexity, i.e. $ \mathcal{O}\big(((K^{2}+2N_{r}^{3})+T_{1}(L+1)^{4.5})log(\epsilon)\big)$. Now, assuming that the algorithm converges in $T_{2}$, the computational complexity of this algorithm is $ \mathcal{O}\big(T_{2}(((K^{2}+2N_{r}^{3})+T_{1}(L+1)^{4.5}))log(\epsilon)\big)$.


\section{Numerical results} \label{numerical results} \vspace{-0.1cm}
In this section, we present numerical results to evaluate the proposed scheme. In our simulations, we assume that a BS with K antennas is located at the center of the polar coordinates. In addition, RIS with L reflecting elements are installed around the BS at the fixed locations to assist in signal transmissions. We assume that the RIS are located on a circle centered at the BS with $\frac{\pi}{4}$ angle and radius of 40.  Also the user and eavesdropper are located at $(30, \beta)$ and $(25, \beta)$, respectively, where $\beta = U[0, \frac{\pi}{2}]$. The noise variances are set as $\sigma_{u}=\sigma_{e}=-105dBm$. We assume that all involved channel coefficients are generated by $H=\sqrt{L_{0}d^{-\varepsilon}}Q$, where $L_{0} = -30 dB$ denotes the path loss at reference distance $d_{0} = 1 m$, $d$ is the link distance, $\varepsilon$ denotes the path loss exponent. The corresponding path
loss exponents is set as $\varepsilon_{Bu}=\varepsilon_{Be}=3.75$ , $\varepsilon_{BI}=\varepsilon_{Iu}=\varepsilon_{Ie}=2.2$, $\varepsilon_{eu}=\varepsilon_{eI}=2.5$ and $Q$ is the Rician components is given by \cite{Cunhua2020,Miao2019}:
\begin{align}  \label{eq.Rician}
	Q=\sqrt{\frac{k}{k+1}}Q^{LoS}+\sqrt{\frac{1}{1+k}}Q^{NLoS} &&
\end{align}
where $k=1$ is the Rician factor, $Q^{LoS}$ is the deterministic Line of Sight (LoS), and the NLoS components $Q^{NLoS}$ are i.i.d. complex Gaussian distributed with zero mean and unit variance. The los component is given by
\begin{align}  \label{eq.Rician los}
	Q^{LoS}=a_{D_{r}}(\nu^{AoA})a_{D_{t}}^{H}(\nu^{AoD}) &&
\end{align}
where $a_{D_{r}}(\nu ^{AoA})$ and $a_{D_{t}}^{H}(\nu^{AoD})$ are defined as
\begin{align}  \label{eq.number antenna}
	a_{D_{r}}(\nu ^{AoA})= [1,e^{2\pi j\times \frac{d}{\lambda}\times sin\nu^{AoA}},...,e^{2\pi j\times \frac{d}{\lambda}\times (D_{r}-1)sin\nu^{AoA}}]^T &&\\
	a_{D_{t}}(\nu ^{AoD})= [1,e^{2\pi j\times \frac{d}{\lambda}\times sin\nu^{AoD}},...,e^{2\pi j\times \frac{d}{\lambda}\times (D_{t}-1)sin\nu^{AoD}}]^T &&
\end{align}
where $D_r$ and $D_t$ are the numbers of antennas or elements at the receiver side and transmitter side,
respectively, $d$ is the antenna separation distance, $\lambda$ is the wavelength, $\nu^{AoD}$ is the angle of departure and $\nu^{AoD}$ is the angle of arrival. It is assumed that $\nu^{AoD}$ and $\nu^{AoA}$ are randomly distributed within $[0, 2\pi]$. For simplicity, we set $\frac{d}{\lambda}=\frac{1}{2}$. The stopping threshold for the alternating optimization algorithms is set as $\epsilon=10^{-3}$. The without jamming scheme is used as a benchmark, where the Eve is deployed with $N_r$ antennas to evaluate the impact of the eavesdropping. In addition, the scheme without RIS is also used as a benchmark, where only the beamformer w is optimize.
\begin{figure}[h!]
	\centering
	\includegraphics[scale=0.9]{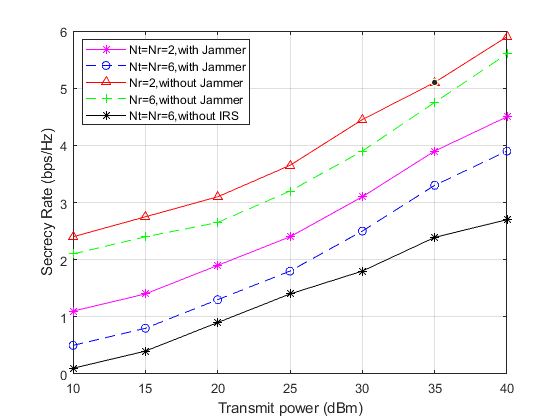}
	\renewcommand{\figurename}{Fig.}
	\vspace{-0.5cm}
	\caption{Achievable Secrecy Rate vs. the Transmit Power, $P_{BS}$.}
	\label{fig.power}
\end{figure}

In \hyperref[fig.power]{Fig.2}, the achievable secrecy rate is plotted against the BS's transmit power for  scenarios both with and without a jammer. The number of antennas ($K$) and reflecting elements in RIS ($L$) are set to 3 and 36, respectively. As we can observe, the secrecy rate increases with an increase in the BS's transmit power in all cases. This increase occurs because as the power of the BS increases, the effect of beamforming on the power received by the user also increases, leading to an increase in the signal-to-noise ratio (SNR) at the user end.
Additionally, increasing the BS's transmit power also increases the power of the received signal in the RIS, which leads to a greater effect of the RIS phase shift optimization on the system's secrecy rate. In other words, the RIS can optimize the phase shift of the reflected signals to enhance the desired signal's power and reduce the interference from the eavesdropper.
Moreover, we can infer from \hyperref[fig.power]{Fig.2} that an increase in the number of eavesdropping antennas results in a decrease in the system's secrecy rate. This is because with more eavesdropping antennas, the eavesdropper can capture more information about the transmitted signal, making it more difficult to maintain secrecy.

\begin{figure}[t!]
	\centering
	\includegraphics[scale=0.9]{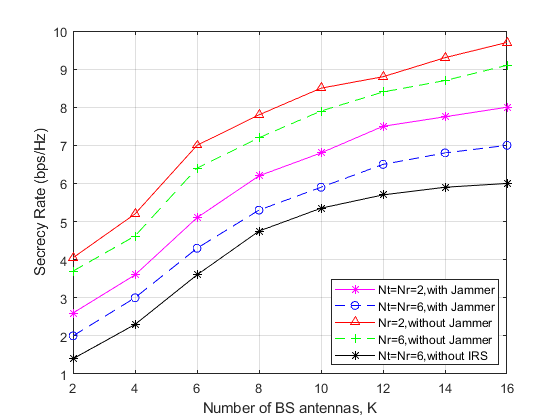}
	\renewcommand{\figurename}{Fig.}
	\vspace{-0.5cm}
	\caption{Achievable Secrecy Rate vs. the number of BS's Antennas, $K$.}
\label{fig.Bs antenna}
\end{figure}
\begin{figure}[h!]
\centering
\includegraphics[scale=0.9]{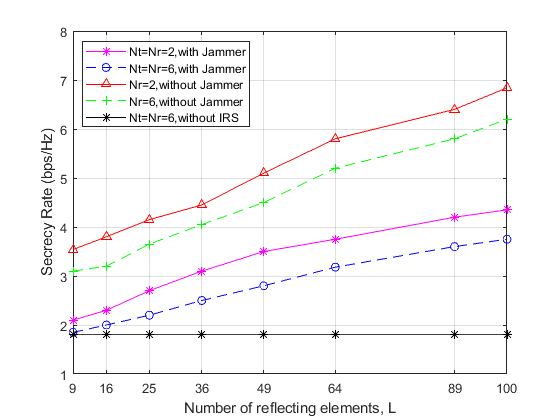}
\renewcommand{\figurename}{Fig.}
\vspace{-0.5cm}
\caption{Achievable Secrecy Rate vs. the number of Reflecting Elements of the RIS, $L$.}
\label{fig.RIS elemans}
\end{figure}
\hyperref[fig.Bs antenna]{Fig.3} illustrates the effect of increasing the number of antennas at the base station on network secrecy rate when the power of the base station and reflecting elements in RIS are set to 30~dBm and 36~dBm, respectively. As shown in the figure, increasing the number of antennas allows for more precise beamforming, resulting in a higher overall secrecy rate. Similarly, an increase in the number of eavesdropper antennas also leads to better beam shaping capabilities, which decreases the secrecy rate. Active attack strategies effectively reduce the achievable secrecy rate by disrupting communication between legitimate transmitter and receiver nodes through jamming signals. Therefore, advanced encryption techniques and physical layer security mechanisms such as artificial noise generation, beamforming, and power control should be deployed to combat these attacks and optimize network security and performance.

\hyperref[fig.RIS elemans]{Fig.4} depicts the secrecy rate versus the number of reflecting elements at the RIS, where $P_{BS} =30 dBm$ and $K=3$. As expected, an increase in the number of RIS elements results in higher secrecy rates, as more phase shift optimization centers become available. Conversely, without reflective elements, the secrecy rate remains constant regardless of the number of such elements added. Interestingly, the gap between the secrecy rates in the presence and absence of a jammer widens with increasing numbers of RIS reflection elements. This is due to the presence of malicious signals sent to the RIS via the jammer, which adversely affects the secrecy rate.

\begin{figure}[t!]
\centering
\includegraphics[scale=0.9]{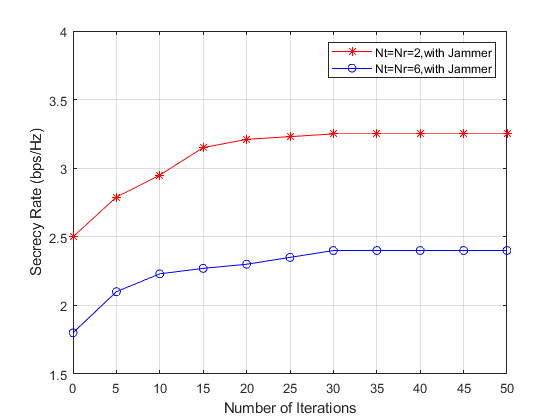}
\renewcommand{\figurename}{Fig.}
\vspace{-0.5cm}
\caption{Convergence of the Proposed BCD Algorithm.}
\label{fig.iteration}
\end{figure}
\hyperref[fig.iteration]{Fig.5} illustrates the convergence of \hyperref[Algorithm Design]{Algorithm 1} in terms of secrecy rate versus the number of iterations, for different numbers of active attacker antennas. The plot shows that the achieved security rate has a non-decreasing trend with increasing number of iterations. Specifically, it can be observed that as the number of iterations increases, the secrecy rate also increases, and \hyperref[Algorithm Design]{Algorithm 1} requires approximately 18 iterations to converge, depending on the number of antennas used. This figure evaluates the effectiveness of the algorithm in solving the problem of maximizing the secrecy rate, clearly demonstrating the increasing trend of the secrecy rate with the number of iterations.


\section{Conclusion}  \label{conclusion}
This paper investigates the security of a RIS-assisted system in the presence of a full-duplex active attacker. The beamforming vector of the base station and the RIS reflecting elements' phases are jointly optimized to maximize the network secrecy rate. To solve the resulting non-convex optimization problem, we propose a novel method based on alternating techniques. Our numerical results demonstrate that the presence of a full-duplex attacker can significantly reduce the network's secrecy rate, especially when the number of antennas is high. The findings indicate that wireless networks using Reconfigurable Intelligent Surface can benefit from an increased rate of secrecy by increasing the number of reflective elements, Base Station power, and the number of antennas. It is crucial to consider these factors when designing secure RIS-assisted systems in the presence of potential attackers.
\bibliography{ref}
\bibliographystyle{IEEEtran}
\end{document}